\documentclass[12pt]{article}

\usepackage{latexsym,enumerate,bm,amsmath,amssymb,color}

\def \corr {\mbox{corr}}
\def \cov {\mbox{cov}}

\usepackage[dvips]{graphicx}

\title{Correction to ``Leverage and volatility feedback effects in high-frequency data''
[J. Financial Econometrics 4 (2006) 353--384]
\footnote{Research partially supported by Spanish grant MTM2007-66632.}}

\author{Amparo Ba\'{\i}llo\footnote{Address for correspondence: Departamento de
Matem\'aticas, Facultad de Ciencias, Universidad Aut\'onoma de Madrid,
28049 Madrid (Spain). Phone: +34 914977641.
e-mail: amparo.baillo@uam.es} \\
{\normalsize \it Universidad Aut\'onoma de Madrid (Spain)}}

\date{}

\begin{document}


\maketitle

\begin{abstract}
\noindent
Bollerslev {\em et al.} (2006) study the cross-covariances for squared returns
under the Heston (1993) stochastic volatility model.
In order to obtain these cross-covariances the authors use an incorrect
expression for the distribution of the squared returns.
Here we will obtain the correct distribution of the squared returns
and check that, under this new distribution,
the result in Appendix A.2 in Bollerslev {\em et al.} (2006) still holds.
\end{abstract}

\newpage

\noindent
{\bf \large 1. Correction to ``Leverage and volatility feedback effects
in high-frequency data''}

Bollerslev {\em et al.} (2006) study the cross-covariances for squared returns
under the Heston (1993) stochastic volatility model
\begin{equation} \label{HestonModel}
\begin{array}{rcl}
dp_t & = &(\mu+cV_t) \, dt +\sqrt{V_t} \, dB_t \\
dV_t & = & \kappa(\theta-V_t) \, dt + \sigma\sqrt{V_t} \, dW_t,
\end{array}
\end{equation}
where $B_t$ and $W_t$ are correlated Brownian motions with $\corr(dB_t,dW_t)=\rho$.
For simplicity, it is assumed that $\mu=c=0$.
If the continuously compounded returns from time $t$ to time
$t+\Delta$ are defined as
$R_{t,t+\Delta} = p_{t+\Delta} - p_t = \int_t^{t+\Delta} \sqrt{V_u} \, dB_u $,
then in Appendix A.2 of Bollerslev {\em et al.} (2006) it is proved that, for $n=0,1,2,\ldots$,
\begin{equation} \label{CrossCovariance}
\cov(R^2_{t+(n-1)\Delta,t+n\Delta},R_{t-\Delta,t})
 = (1-\kappa a_\Delta)^{n-1} \rho \, \sigma \, \theta \, a_\Delta^2 ,
\end{equation}
where $a_\Delta = (1-e^{-\kappa\Delta})/\kappa$.

In order to obtain (\ref{CrossCovariance}) these authors use the following distribution of the
squared returns
\begin{equation} \label{OldSquaredReturns}
R^2_{t,t+\Delta} = 2 \int_t^{t+\Delta} R_{u,u+\Delta} \, \sqrt{V_u} \, dB_u
+ \int_t^{t+\Delta} V_u \, du.
\end{equation}
Observe, however, that $R^2_{t,t+\Delta} = (p_{t+\Delta} - p_t)^2 $ cannot
depend on the values of $R_{u,u+\Delta}$ with $u\in[t,t+\Delta]$, that is, on returns
which are posterior to $t+\Delta$. Here we will obtain the correct expression for
the distribution of the squared returns and check that, under this new distribution,
result (\ref{CrossCovariance}) in Bollerslev {\em et al.} (2006) still holds.

In order to obtain the distribution of the squared returns observe that
$R^2_{t,t+\Delta} = p_{t+\Delta}^2 + p_t^2 -2\,p_t\,p_{t+\Delta}$. By It{\^o}'s Lemma we have that
\begin{equation} \label{SquaredLogPrice}
p_{t+\Delta}^2 = p_t^2 + 2\int_t^{t+\Delta} p_u \, \sqrt{V_u} \, dB_u
 + \int_{t}^{t+\Delta} V_u \, du.
\end{equation}
Using (\ref{HestonModel}) and (\ref{SquaredLogPrice}) we have that
\begin{equation} \label{DistributionSquaredReturns}
R^2_{t,t+\Delta} = 2 \int_t^{t+\Delta} R_{t,u} \, \sqrt{V_u} \, dB_u
+ \int_t^{t+\Delta} V_u \, du.
\end{equation}
The only difference between (\ref{OldSquaredReturns}) and
(\ref{DistributionSquaredReturns}) is the first integral.
This is why a large part of the proof of (\ref{CrossCovariance}) in
Bollerslev {\em et al.} (2006) remains valid.
More concretely, it is still true that
$$
\cov(R^2_{t+(n-1)\Delta,t+n\Delta},R_{t-\Delta,t})
 = E \left( \int_{t+(n-1)\Delta}^{t+n\Delta} V_u \, du
 , \int _{t-\Delta}^t \sqrt{V_u} \, dB_u \right).
$$
To show this it may easily be checked that
$ E \left( \int_{t+(n-1)\Delta}^{t+n\Delta} R_{t+(n-1)\Delta,u} \, \sqrt{V_u} \, dB_u
  \, | \, F_{t+(n-1)\Delta} \right) = 0 $.


\

\noindent
{\bfseries References}

\begin{description}
\item Bollerslev, T., Litvinova, J. and Tauchen, G. (2006).
Leverage and volatility feedback effects in high-frequency data.
{\em Journal of Financial Econometrics}, 4, 353--384.
\item Heston, S. L. (1993). A closed-form solution for options with stochastic
volatility with applications to bond and currency options. {\em The Review of
Financial Studies}, 6, 327--343.
\end{description}

\end{document}